\documentclass[twocolumn,aps,superscriptaddress,amsfonts,floatfix]{revtex4}
\usepackage{epsfig}
\usepackage[T1]{fontenc}
\def\openone{\leavevmode\hbox{\small1 \normalsize \kern-.64em1}}
\begin{document}
\title{Quantum Wire as Open System}
\author{Marcin Wie\'sniak}
\affiliation{Centre for Quantum Technologies, National University of Singapore, 3 Science Drive 2, Singapore 117543, Singapore}
\affiliation{Department of Physics, National University of Singapore, Singapore 117542, Singapore}
%\affiliation{Instytut Fizyki Teoretycznej i Astrofizyki, Uniwersytet Gda\'nski, PL-80952 Gda\'nsk, Poland.}
\begin{abstract}
The faithful exchange of quantum information will soon become one of the challenges of the emerging quantum information technology. One of the possible solutions is to transfer a superposition through a chain of properly coupled spins. Such a system is called a quantum wire. We discuss the transfer in a quantum wire \cite{christ,niko1,niko2}, when the process of thermalization of the state takes place together with the free evolution. We investigate which encoding scheme is more faithful in certain thermal conditions.
\end{abstract}
\maketitle
%How can one imagine a quantum computer? Logically, it would be distinguished by taking the advantage of the superposition principle \cite{shor,deutsch,grover}. But from the more technical point of view, it would be not very different from present, classical machines. It would contain various modules. Each of this part would be responsible for different functions and they would need to exchange (quantum) information with one another.
\section{Introduction}
The emerging quantum information technology elevates faithful exchange of quantum states to the rank of an important problem. 
Quantum mechanics allows to send information even without any flow of energy. One possibility is the teleporation \cite{teleportation}. It is a procedure, in which one participant performs a measurement on a particle in an unknown state and one more particle to announce its result. Having received this message, the other party can reproduce the teleported state on his quantum system (the original is destroyed). This scheme relies on the resource of quantum entanglement, which might be difficult to produce. Another option is to use a line of mutually coupled spins$-\frac{1}{2}$ representing quantum bits (qubits). This coupling should be a natural interaction, rather than a sequence of on-demand SWAP gates, to minimize the amount of the external control. A uniform Heisenberg chain with uniform nearest neighbor coupling, like in Sr$_2$CuO$_3$ \cite{sr2cuo3}, is not efficient as a quantum wire. Its free evolution is only quasi-periodic, but not periodic. As a result, information encoded at one end of the chain will spread over the whole system, never to refocus at the other. This problem was addressed by e.g. Bose \cite{bose}, and subsequently by Burgarth, Giovannetti, and Bose \cite{bgb}.  In the latter article the receiver performs at certain times the partial SWAP operation. This time-dependent action gradually transfers the encoded message to an auxiliary qubit. This solution can provide an arbitrarily close to perfect transmission, but requires a precise control and a relatively large computing power, as the partial swap should be in correspondence to the free evolution of the state.

An alternative system was proposed by Christandl {\em et al.} \cite{christ}, and independetly by Nikolopoulus \cite{niko1,niko2}. They designed a spin chain with modulated interspin coupling constants and a periodic free evolution. Whichever state from the certain subspace (the one, in which exactly one spin-$\frac{1}2$ is pointing up, rather than down, giving the total $z$-magetization equal to $N/2-1$, herein called one quasi-particle subspace or OQS) is created at the beginning of evolution, it is mirrored with respect to the middle of the chain after one half of the period. This scheme was then attempted to be refined by Kay \cite{kay}, who investigated the role of non-nearest neighbor interactions.

This feature enables to encode a logical qubit in a few different ways. Following e.g. \cite{christ} one can (a) encode it in the subspace of the state in which all spins are aligned with respect to the $z$-axis and one of the states spanning OQS, in particular, the one in which the first spin of the chain is anti-parallel to the rest. Another possibility is found in \cite{danielbose}, where the Authors propose (b) using the dual rail, that is two independent chains. An excitation in one of them will encode the logical $|0\rangle$, in the other -- the logical $|1\rangle$. %One may also think of realizing this code with a single chain (c). One can use the anti-alignment of the first qubit to encode $|0\rangle$, whereas $|1\rangle$ shall be represented by a flip of the second spin in the chain. 
The second scheme requires not only a writing head,
\begin{eqnarray}
|0\rangle_{in}|00\rangle_{ab}\rightarrow|0\rangle_{in}|10\rangle_{ab},\nonumber\\
|1\rangle_{in}|00\rangle_{ab}\rightarrow|0\rangle_{in}|01\rangle_{ab},
\end{eqnarray}
where the subscript $in$ denotes the input data qubit and $a,b$ are two initial spins, but also a read-out head: 
\begin{eqnarray}
&|10\rangle_{cd}|0\rangle_{out}|0\rangle_h\rightarrow|00\rangle_{cd}|0\rangle_{out}|0\rangle_h,&\nonumber\\
&|01\rangle_{cd}|0\rangle_{out}|0\rangle_h\rightarrow|00\rangle_{cd}|1\rangle_{out}|0\rangle_h,&\nonumber
\end{eqnarray}
with $c,d$ being the final qubits, $out$--the target qubit, and $h$--the head ancilla. In presence of decoherence, the head should recognize the message as destroyed and set the target qubit state maximally mixed when the two final spins are co-aligned with respect to the $z$-axis:
\begin{eqnarray}
|00\rangle_{cd}|0\rangle_{out}|0\rangle_h\rightarrow\frac{1}{\sqrt{2}}|01\rangle_{cd}(|0\rangle_{out}|1\rangle_h+|1\rangle_{out}|2\rangle_h)&\nonumber\\
|11\rangle_{cd}|0\rangle_{out}|0\rangle_h\rightarrow\frac{1}{\sqrt{2}}|10\rangle_{cd}(|0\rangle_{out}|1\rangle_h-|1\rangle_{out}|2\rangle_h).\nonumber
\end{eqnarray}

Also, we can propose yet another encoding (c): $|0\rangle$ is physically represented by a flip of the first spin, $|10...0\rangle$, whereas logical $|1\rangle$ is associated to a flip of the second qubit, $|01...0\rangle$. The code utilizes the same heads as (b), but only one wire.

The aim of this report is to study the robustness of encodings (a) and (c) against the interaction of the wire with an external heat bath. The dual-rail scheme is not consiedered, as we assume that doubling the number of spins makes the system more sensitive to thermalization. The problem of decoherence of a quantum wire was already discussed in a number of papers \cite{chiny,danielbose2,kay1,exp3}. However, in these articles the Authors have considered errors addressed to individual spins. Herein, we rather consider coupling with the heat bath through the eigenstates of the Hamiltonian, which (except for some cases) leads to the Gibbs state $\rho=1/Z\exp(-\beta H)$. Here, $\beta=1/(\kappa T)$, $T$ stands for the temperature of the heat bath, $\kappa$ denotes the Boltzmann constant, $H$ is the Hamiltonian and $Z=Tr\exp(-\beta H)$ is the partition function. The motivation for this approach could be that in some systems, the wavelengths corresponding to transitions due to decoherence are much  longer than the physical length of a wire. In the regime of long waves it is difficult to consider localized errors.

\section{Formulation of Problem}
We start with the Hamiltonian similar to the one given in \cite{christ,niko1,niko2},
\begin{equation}
\label{hamiltonian}
H=J\sum_{i=1}^{N-1}\sqrt{i(N-i)}(\sigma^x_{i}\sigma^x_{i+1}+\sigma^y_{i}\sigma^y_{i+1}).
\end{equation}
$\sigma^x$ and $\sigma^y$ are the $x$- and $y$- components of the Pauli matrices vector (together with $\sigma^z$), the subscript denotes the spin being acted on, and $J=\pi$ is chosen such that the period of the free evolution is 1. It is necessary to adopt the convention that the operator acting on the spin with the lower number stands in front of the other.

The only difference is that while most Authors consider the Heisenberg interaction of the type $\vec{\sigma_{i}}\cdot\vec{\sigma_{i}}$, we rather consider the exchange terms, $\sigma^x_{i}\sigma^x_{j}+\sigma^y_{i}\sigma^y_{j}$. This is not relevant for the perfect state transfer, nevertheless, we can take the advantage of the exact solvablity of 1-D $xy$ spin$-\frac{1}{2}$ models, first shown by Katsura \cite{katsura}. Which interaction actually occurs is a characteristic of a used physical system.

The first step to diagonalize (\ref{hamiltonian}) is to perform the Jordan-Wigner transformation. %We define the operators $\sigma_\pm^{[i]}=\frac{\sigma_x^{[x]}\pm i\sigma_y^{[y]}}{2}$. 
By choosing
\begin{eqnarray}
\label{jordanwigner1}
&\hat{a}_{i}=\frac{1}{2}\left(\prod_{j=1}^{i-1}\sigma^z_{j}\right)\left(\sigma^x_{j}+i\sigma^y_{j}\right),\\
\label{jordanwigner2}
&\hat{a}^{\dagger}_i=\frac{1}{2}\left(\prod_{j=1}^{i-1}\sigma^z_{j}\right)\left(\sigma^x_{j}-i\sigma^y_{j}\right),
\end{eqnarray}
one constitutes the canonical anti-commutation relations (CAR),
$\{\hat{a}_{i},\hat{a}^{\dagger}_j\}=\delta_{i,j},$ $\{\hat{a}_{i},\hat{a}_{j}\}=0$,
and gets $\sigma^{x}_{i}\sigma^{x}_{i+1}+\sigma^{y}_{i}\sigma^{y}_{i+1}=\hat{a}^{\dagger }_{i+1}\hat{a}_{i}+\hat{a}^{\dagger}_{i}\hat{a}_{i+1}$. Hence (\ref{hamiltonian}) can be expressed in terms of modes of non-interacting quasi-fermions:
\begin{equation}
\label{fermionh}
H=\hat{A}^\dagger H_{OQS}\hat{A},
\end{equation}
with $\hat{A}=(\hat{a}_1,...,\hat{a}_N)^T$ and $H_{OQS}$ representing the projection of the Hamiltonian onto OQS. It is now clear, that the state $|0\rangle^{\otimes N}=|\Omega\rangle$ is the quasi-fermionic vacuum and OQS is spanned by one quasi-fermion states.

In some physical implementaions of the wire, e.g., those based on Josephson junctions (see e.g. \cite{exp1,exp2}), we can also equip each quasi-fermion with some energy of its creation (expressed as frequency $\omega$) by adding the magnetic term $\frac{\omega}{2}\sum_{i=1}^N(1-\sigma^z_{i})$ ($\hbar=1$) to the Hamiltonian. This corresponds to replacing $H_{OQS}$ in (\ref{fermionh}) with  $ H_{OQS}+\omega\openone_{N\times N}$. Herein, we will work in the limit of $\omega \gg JN$. The energy of the state is practically determined by the number of quasi-fermions. On the other hand, this will cause fast oscillations of the relative phase between, e.g., the vacuum state and a state of one quasi-fermion. Thus we choose $\omega/(2\pi)$ as a large, even integer. The phase will then oscillate an integer number of times within half of the period and these oscillations can be neglected.

The second major step in the solution of Katsura is a unitary transformation of the quasi-fermionic modes, $\hat{c}_{i}=\sum_{j=1}^Nb_{ij}\hat{a}_{j}$, which preserves CAR. While the Fourier transform is used in \cite{katsura}, in our case the particular transformation is the one, which diagonalizes $H_{OQS}$. Note that elements of $\{b_{ij}\}$ can be always chosen real. Keeping in mind our choice of the period, we get	
\begin{equation}
\label{diagon}
H=\sum_{i=1}^N(E_{i}+\omega)\hat{c}^{\dagger}_i \hat{c}_{i},
\end{equation} 
where $E_{i}=-(N-1)\pi,-(N-3)\pi,...,(N-1)\pi$ for $N$ even and $E_{i}=-(N-1)\pi,...,0,...,(N-1)\pi$ for $N$ odd. 

The Reader will notice that in scheme (a) for even $N$ the transfered state experiences a phase flip. One may either change $\omega$ by $\pi$ or let the receiver correct this deformation manually.

The transfer procedure is as follows. On demand, the state of the wire is brought to the vacuum state. For $\omega>(N-1)\pi$ this can be done by cooling the system. Subsequently, the transmitee state is is encoded to the first (a) or the first two (c) spins of the chain. The receiver waits half of the period and decodes the message at his end. On top of the free evolution, as we assume, the wire is coupled to an external heat bath. This coupling occurs through the Hamiltonian eigenmodes and shall be seen as annihilation or creation of quasi-fermions in the wire, accompanied by annihilating or creating bosons in the external field. Let us take the probability per unit time of loosing $i$th quasi-fermion as $\gamma_i$ and, according to the quantum detailed balance condition \cite{quantumbc}, the probability per unit time of the inverse process reads $\gamma_i \exp(-\beta(\omega+E_i))$. Hence the master equation governing the evolution \cite{mastereq} is
\begin{eqnarray}
\label{master}
\frac{\partial\rho}{\partial t}=&&L(\rho)\nonumber\\
=&&-i[H,\rho]\nonumber\\
+&&\sum_{i=1}^N\frac{\gamma_i}{2}\left(2\hat{c}_{i}\rho\hat{c}^{\dagger}_i-\hat{c}^{\dagger}_i\hat{c}_{i}\rho-\rho\hat{c}^{\dagger}_i\hat{c}_{i}\right)\nonumber\\
+&&\sum_{i=1}^N\frac{\gamma_i}2\exp(-\beta(\omega+E_i))\left(2\hat{c}^{\dagger}_i\rho\hat{c}_{i}-\hat{c}_{i}\hat{c}^{\dagger}_i\rho-\rho\hat{c}_i\hat{c}^{\dagger}_i\right).\nonumber\\
\end{eqnarray}

It is noteworthy that, since the action of $L$ involves the energy states, one can neglect the free evolution and instead study the fidelity of the image of the initial state.

Decoherence will most generally act on the state as the following map:
\begin{equation}
%\left(\begin{array}{c}\openone\\ \sigma^x\\ \sigma^y\\ \sigma^z\end{array}\right)_{in}\rightarrow 
\left(\begin{array}{c}\openone\\ \sigma^x\\ \sigma^y\\ \sigma^z\end{array}\right)_{out}=\left(\begin{array}{cccc} 1&0&0&0\\ 0&\lambda_{xx}&0&\lambda_{xz}\\ 0&0&\lambda_{yy}&\lambda_{yz}\\ \lambda_{z0}&\lambda_{zx}&\lambda_{zy}&\lambda_{zz}\end{array}\right)\left(\begin{array}{c}\openone\\ \sigma^x\\ \sigma^y\\ \sigma^z\end{array}\right)_{in}.
\label{mapa}
\end{equation}
The quantity of our interest is the average fidelity,  $F=\int|\langle|\psi|_{in}\rho_{out}|\psi\rangle_{in}|^2d\mu(|\psi\rangle_{in})$ ($\int d \mu(|\psi\rangle_{in})=1$). From (\ref{mapa}) we find it to be $F=\frac{1}{2}(1+(\lambda_{xx}+\lambda_{yy}+\lambda_{zz})/3)$. 
There are two important thresholds on mean fidelity. If it equals $\frac{1}{2}$, no effort was taken for the transmission. If it is not above $\frac{2}{3}$, no quantum information was transfered. Equally well, the sender could have performed a measurement on his qubit and announce the result. The receiver could then have aligned his spin respectively to the message.

\section{Analytical Results}
In this section we investigate effects of decoherence in the limit of weak coupling, $\gamma_i\rightarrow 0$. Then, at half of the period of the free evolution, when the transfer should have occurred, the state is close to:
\begin{eqnarray}
\rho_\frac{1}{2}&=&|\psi_0\rangle\langle\psi_0|+\frac{1}{2}L(|\psi_0\rangle\langle\psi_0|)\nonumber\\
&=&|\psi_0\rangle\langle\psi_0|\nonumber\\
&+&\sum_{i=1}^N\frac{\gamma_i}2(\hat{c}_i|\psi_0\rangle\langle\psi_0|\hat{c}_i^\dagger\nonumber\\
&-&\frac{1}{2}\hat{c}_i^\dagger\hat{c}_i|\psi_0\rangle\langle\psi_0|-\frac{1}{2}|\psi_0\rangle\langle\psi_0|\hat{c}_i^\dagger\hat{c}_i)\nonumber\\
&+&\sum_{i=1}^N\frac{\gamma_i}2\exp(-\beta(E_{i}+\omega))(\hat{c}_i^\dagger|\psi_0\rangle\langle\psi_0|\hat{c}_i\nonumber\\
&&-\frac{1}{2}\hat{c}_i\hat{c}_i^\dagger|\psi_0\rangle\langle\psi_0|-\frac{1}{2}|\psi_0\rangle\langle\psi_0|\hat{c}_i\hat{c}_i^\dagger,\nonumber)\\
\label{mastereq1}
\end{eqnarray}
where $|\psi_0\rangle$ is the initial state. 

In order to extract the elements of the map, we send the Pauli operators, rather than states. Encoding (a) involves the following operators to construct the initial states:
\begin{eqnarray}
\openone^{(a)}=&&|\Omega\rangle\langle\Omega|+\hat{a}_1^\dagger|\Omega\rangle\langle \Omega|\hat{a}_1,\\
\sigma^{z(a)}=&&|\Omega\rangle\langle\Omega|-\hat{a}_1^\dagger|\Omega\rangle\langle\Omega|\hat{a}_1,\\
\sigma^{+(a)}=&&|\Omega\rangle\langle\Omega|\hat{a}_1
\end{eqnarray}
($\sigma^\pm=\frac{1}{2}(\sigma^x\pm i\sigma^y),(\sigma^+)^\dagger=\sigma^-$), while in encoding (c) we have
\begin{eqnarray}
\openone^{(c)}=&&\hat{a}_1^\dagger|\omega\rangle\langle\Omega|\hat{a}_1+\hat{a}_2^\dagger|\Omega\rangle\langle\Omega|\hat{a}_2,\\
\sigma^{z(c)}=&&\hat{a}_1^\dagger|\Omega\rangle\langle\Omega|\hat{a}_1-\hat{a}_2^\dagger|\Omega\rangle\langle\Omega|\hat{a}_2,\\
\sigma^{-(c)}=&&\hat{a}_1^\dagger|\Omega\rangle\langle\Omega|\hat{a}_2.
\end{eqnarray}
Consequently the average fidelities read
\begin{eqnarray}
F^{(a)}=&&1\nonumber\\
+&&\frac{1}{24}\underbrace{Tr(\hat{a}_1^\dagger L(\sigma^{z(a)})\hat{a}_1-\hat{a}_1L(\sigma^{z(a)})\hat{a}_1^\dagger)}_{=2F^{(a)}_z}\nonumber\\
+&&\frac{1}{6}\underbrace{Tr\hat{a}_1L(\sigma^{+(a)})}_{=F^{(a)}_{xy}}\\
F^{(c)}=&&1\nonumber\\
+&&\frac{1}{24}\underbrace{Tr(\hat{a}_1PL(\sigma^{z(c)})P\hat{a}_1^\dagger-\hat{a}_2PL(\sigma^{z(c)})P\hat{a}_2^\dagger}_{=2F^{(c)}_z}\nonumber\\
+&&\frac{1}{6}\underbrace{Tr\hat{a}_2^\dagger PL(\sigma^{+(c)})P\hat{a}_1}_{=\frac{1}{2}(F^{(c)}_{x}+F^{(c)}_{y})}.
\end{eqnarray}
$P=\hat{a}_1\hat{a}_1^\dagger\hat{a}_2^\dagger\hat{a}_2+\hat{a}_1^\dagger\hat{a}_1\hat{a}_2\hat{a}_2^\dagger$ is the projector ruling out the case of the two spins being aligned in the $z$ direction.

Plugging (\ref{master}) to the above we get that at infinite temperature ($\beta=0$) $F^{(c)}_x=F^{(c)}_{y}=F^{(c)}_{z}=-\sum_{i=1}^N\gamma_i(b_{i1}^2+b_{i2}^2)$, $F^{(a)}_z=-\sum_{i=1}^N2\gamma_ib_{i1}^2$, and $F^{(a)}_{xy}=\sum_{i=1}^N\gamma_i(b_{1i}^2-2)$. Thus, independently of coupling constants, the encoding (c) is more robust against the weak thermalization than (a). In particular, when $\forall_i\gamma_i=\Gamma$ One gets $F^{(c)}=1-\frac{\Gamma}{2}$ and $F^{(a)}=1-\Gamma\frac{N}{3}$. For long chains at high temperature, decoherence quickly removes the quantum features of the transferred information in  scheme (a).

Considering the limit of $T\rightarrow 0$ and the case of $\omega>(N-1)\pi$ in the same fashion, one gets that $F^{(a)}=1-\frac{1}{4}\sum_{i=1}^N\gamma_i$ and $F^{(c)}=1-\frac{1}{2}\sum_{i=1}^N\gamma_ib_{1i}^2$. In fact, in this case (\ref{master}) is exactly solvable and gives $F^{(a)}(t)=\frac{1}{2}(1+\sum_{i=1}^N\exp(-\gamma_it/2))$ and $F^{(c)}(t)=\frac{1}{2}(1+\sum_{i=1}^N\exp(-\gamma_it)$. This is because all states are irreversibly transformed into the vacuum state, which alone cannot transfer any information.

These two results show that the advantage of one encoding over the other depends on thermal conditions. At low temperatures, fidelities in both schemes decay exponentially, but in (c) the decay is faster. At high temperatures, on contrary, decoherence in scheme (a) acts mainly on coherences, with the strength proportional to $N$. Note that the high temperature behavior discussed above is true, independently of $\omega$ and $\gamma_i$s 

Let us now pass to the detailed discussion on the scheme (c). Let us assume $\omega\rightarrow\infty$. We thus take the rescaled inverse temperature, $\beta'=\omega\beta$. We also take $\forall_i\gamma_i=\gamma$. It was confirmed numerically that 
\begin{equation}
\label{expon21}
F^{(c)}(\Gamma,t)\approx\frac{1}{2}(1-p_{ap}(\Gamma,t)\exp(-a_1\exp(-\beta')\Gamma^2t^2))
\end{equation}
and
\begin{equation}
\label{expon22}
p_{ap}(\Gamma,t)\approx p_1(\Gamma,t)\exp(-a_2\exp(-\beta')\Gamma^2t^2)).
\end{equation}
Here, when we switch off the free evolution, $p_{ap}(\Gamma,t)$ is the probability that the two spins are anti-parallel with respect to the $z$-axis. $p_1(\Gamma,t)$ is the probability that the whole system contains exactly one quasi-fermion. Hence $p_1(\Gamma,t)$ can be found from the classical master equation for populations.  

Numerics show that $a_1=2$ and $a_2=N-2$. This allows to interpret equations (\ref{expon21},\ref{expon22}). The factor $\exp(-\beta')\Gamma^2$ is a product of per-unit-time probabilities of subsequent annihilating and creating (or the otherwise) a quasi-particle. This can be seen as a collision of the quasi-fermion with field boson; the number of particles (and in this special case, the energy of the wire) is conserved, but quantum numbers change. The value of $a_2$ shows that (\ref{expon22}) is the effect of annihilation of the quasi-fermion in 2-dimensional subspace of OQS carrying the information, and its creation in the remaining, irrelevant $(N-2)$ dimensions of OQS. Relation (\ref{expon21}) should then correspond to swaps within the essential 2D space. Note that $a_1$ is expected to be 1, rather than 2, as compared to $a_2$. This is because the decohered state not only looses its overlap with its original, but also has an increased scalar product with the other state.

A similar, but more complex behavior of $F^{(c)}$ with respect to $p_1$ can be observed even if $\omega$ is not significantly larger than $\max E_i$ or the absorbsion/emission rates are not uniform. In either case, the decay is expected to be rather of the type $F^{(c)}\approx\frac{1}{2}\left(1+p_1\sum_ik_i\exp(-l_it^2)\right)$, with $\forall_i l_i>0$, $\sum_ik_i=1$ and $l_i$'s being exponentially decaying functions of $\beta$. The reason for this complication is that now every  collision involving a given pair of quasi-fermions happens at its own rate. However, this tells us that the fidelity of transfer is at least upper bounded by $(1+p_1)/2$, which, in general, can be obtained from the solution of the Pauli equation for diagonal elements of $\rho$.
\section{Numerical results}
In the following section we will discuss the competition between the encoding strategies on few potentially interesting examples. 

The first case is of $\omega\ll 1/\tau$ and $\forall_{i}\gamma_i=\Gamma$. This allows us to introduce a new scale for temperature, $\beta'=\omega\beta$. The numerical simulation was done for $N=6$ (Figure 1) and $N=7$ (Figure 2). Te scheme (c) is more faithful in the whole region plotted in Figures 1 and 2. While the region of $F^{(c)}>2/3$ (left from solid lines) seems to be stable with $N$, the part ofthe plot with of $F^{(a)}>2/3$ (left from dotted lines) is larger for $N=6$ and $N=7$. Consequently, the grey part of the plot, where we have $F^{(c)}>F^{(a)}$, shrinks. 

\begin{figure}
\center
\includegraphics[width=6cm]{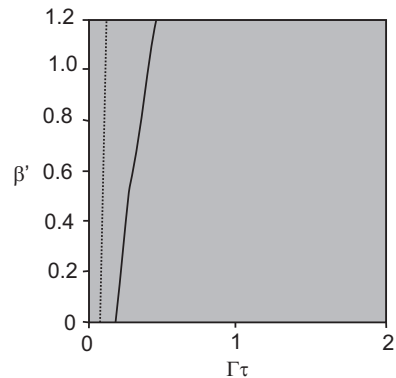}
\caption{Fidelity thresholds for encodings (a) and (c) for $N=6$ and $\omega\rangle\infty$ as functions of $\Gamma\tau$ and $\beta'$. In all Figures, $F^{(c)}>2/3$ occurs on the left from the solid line, $F^{(a)}>2/3$ is left from the dotted curve, and the region of $F_{(c)}>F_{(a)}$ is colored gray.}
\end{figure}

\begin{figure}
\center
\includegraphics[width=6cm]{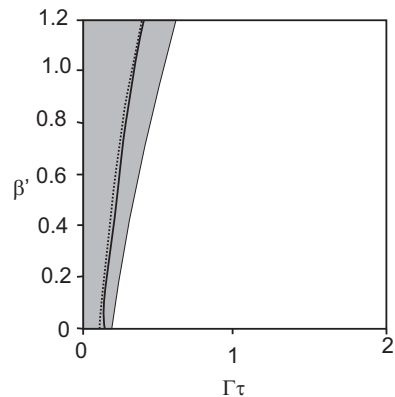}
\caption{Fidelity thresholds for encodings (a) and (c) for $N=6$ and $\omega\rightarrow\infty$ as functions of $\Gamma\tau$ and $\beta'$.}
\end{figure}

\begin{figure}
\center
\includegraphics[width=6cm]{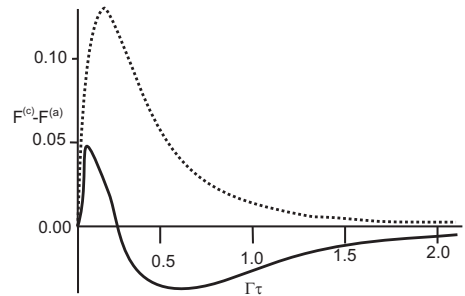}
\caption{Plots of $F_{(c)}-F_{(a)}$ at $\beta'=0$ for $N=6$ (dotted) and $N=7$ (solid).}
\end{figure}

As $\omega$ lessens, the wire might experience a series of quantum phase transitions. As can be seen from eqn. (\ref{diagon}), the first such transitions for the considered length of $N=6$ happens at $\omega=5\pi$. It is a change of the ground state from the quasi-fermionic vacuum to a state from $OQS$. This phase transition affects the process of thermalization at finite temperatures. The Figure 4 presents curves $F_{(a)}=2/3$, $F_{(c)}=2/3$ and the region of $F_{(c)}>F_{(a)}$ for $\omega=5.01\pi$. The coupling constants have been taken $\gamma_i=\gamma(E_i+\omega)^2$. The quadratic dependence of $\gamma_i$s on the energy would follow from the fact that the coupling strength should be proportional to the density of modes of the given frequency.

\begin{figure}
\center
\includegraphics[width=6cm]{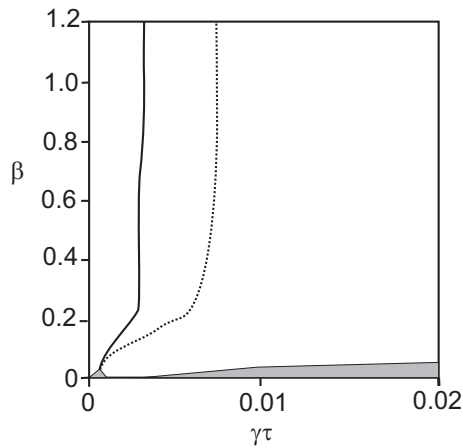}
\caption{Fidelity thresholds for encodings (a) and (c) for $N=6$ and $\omega=5.01\pi$ as functions of $\gamma\tau$ and $\beta$.}
\end{figure}

We observe that scheme (c) is advantageous over (a) only at very low temperatures. The advantage soon vanishes to return for higher $\gamma\tau$, but for these values, neither of the schemes allows to send the state coherently.

The next value of $\omega$, which is potentially interesting, is $4\pi$. As we have argued in the previous paragraph. OQS plays a special role in scheme (c). For $3\pi<\omega<5\pi$, the ground state lies in that subspace, hence one could expect a higher efficiency of the two-spin encoding. For the mode, for which $\omega+E_i<0$, rhe roles of the absorbsion and the emission are interchanged. Again, $\gamma_i=\gamma(\omega+E_i)^2$. Figure 5 shows that, as  compared to the previous case, (c) is more robust against decoherence only at high temperatures. However, at $\beta=0$ (a) is never more optimal.

\begin{figure}
\center
\includegraphics[width=6cm]{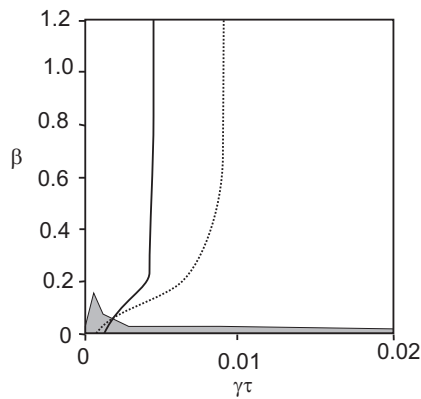}
\caption{Fidelity thresholds for encodings (a) and (c) for $N=6$ and $\omega=4\pi$ as functions of $\gamma\tau$ and $\beta$.}
\end{figure}

Finally, we investigate the case of $\omega=0$ an $\gamma_i$s as above. This seems to be a natural regime for spin-based implementation of the wire. We would like to make a note, that, in general, (\ref{master}) might be not sufficient as a model of decoherence. For $N$ odd, there exist a mode of the total energy 0, which is hence not coupled to the environment. In such a case, one should also include processes involving more than one quasi-fermion. These events can be taken neglibly unlikely as compared to the most probable one quasi-fermion transitions. Within our model, we are still able to make some estimates with calculations for $N=6$. Figure 6 suggests that the scheme $(c)$ provides a higher fidelity than $(c)$, even at low temperatures. $F_{(a)}>F_{(c)}$ was never observed at $\beta=0$.

\begin{figure}
\center
\includegraphics[width=6cm]{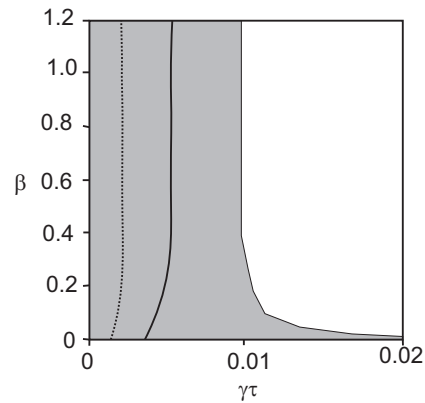}
\caption{Fidelity thresholds for encodings (a) and (c) for $N=6$ and $\omega=0$ as functions of $\gamma\tau$ and $\beta$.}
\end{figure}

\section{Summary}
In this article, we have addressed the problem of the interaction of a quantum wire with an external heat bath during the state transfer. We have taken a simple case of $xx$ inter-spin coupling and the change of quasi-fermions from the wire into bosons from the external field, and vice verse, as an elementary process of decoherence. We have compared two strategies of encoding: (a), when two logical states are encoded into two states of one physical qubit, and (c), when each logical state corresponds to a flip of different spin. The dual-rail scenario (b) has been discarded, basing on the assumption that the overall effect of thermalization is stronger on the bigger infrastructure.

Our thermalization model could be suitable for chains, which couple to waves longer than the physical length of the wire. This is the case, when spin-spin interactions are weak. Long wavelengths allow to address the constituents of the chain collectively, rather than each of them individually. They are also responsible for the high susceptibility of the system to temperature. Hence in realistic implementations, the high $T$ regime will be of a great relevance.

In general, neither of the schemes is more optimal than the other. We have proven that at $\beta=0$ and for some time (c) is more faithful than (a), regardless of values of $\gamma_i$s. We have also shown numerically, that the latter scheme is more robust even at lower temperatures in two interesting cases of $\omega\rightarrow\infty$ and $\omega=0$. One should point out, however, that for large $\omega$ the advantage of (c) over (a) was considerably smaller for the longer chain. 

A complicated competition between the encodings originates from two different behaviors of the encoded state. In the single spin encoding, the off-diagonal elements of the state, which express the superpositions of different number of quasi-particles, are most affected. In case of two-spin encoding, we deal with the migration of particles, and their collisions. Initially, collisions are highly probable, As time (or $\gamma_i$s) grows, the decay of fidelity due to collisions becomes dominant. Eventually, scheme (a) turns out to be more faithful.

The Author would like to acknowledge D. Kaszlikowski, R.W. Chhajlany, M. Piani, P. Badzi\k{a}g, and A. Kay %, and D. Burgharth
 for useful and teaching discussions. The Author thanks R. Alicki for an inspiration. 

This work is a part of EU 6FP programmes QAP. The Author receives the START scholarship from the Foundation for Polish Science (FNP). This work is supported by the National Research  
Foundation and Ministry of Education, Singapore. 	

\end{document}